\documentstyle[aps]{revtex}
\begin{document} 
\title{The $n\to\infty$ limit of $O(n)$ models on graphs}
\author{Raffaella Burioni\footnote{E-mail:
burioni@pr.infn.it}, Davide Cassi\footnote{E-mail: cassi@pr.infn.it}}
\address{ Dipartimento di Fisica, INFM and INFN, Universit\`a di Parma,
          Viale delle Scienze, 43100 Parma, Italy 
              }
\author{Claudio Destri\footnote{E-mail:
destri@mi.infn.it}}
\address{ Dipartimento di Fisica and INFN, Universit\`a di Milano,
          Via Celoria, 16  20133 Milano, Italy 
              }
\date{\today}
\maketitle
\begin{abstract} 
Thirty years ago, Stanley showed that an $O(n)$ spin model on a lattice  
tends to a spherical model as $n\to\infty$. This means that
at any temperature the corresponding free energies coincide.
This fundamental result, providing the basis for more detailed studies
of continuous symmetry spin models, is no longer valid on more general
discrete structures lacking of translation invariance, i.e. on graphs.
However only the singular parts of the free energies determine the
critical behavior of the two statistical models. 
Here we show that such singular
parts still coincide even on general graphs in the thermodynamic limit.
This implies that the critical exponents of $O(n)$ models on graphs
for $n\to\infty$ tend to the spherical ones and therefore they only 
depend on the graph spectral dimension. 
\end{abstract}


The study of spin models on non crystalline structures is
an intriguing and complex problem in statistical mechanics. 
This is fundamentally due to the lack of translational invariance 
and of a natural definition for the system dimensionality. 
The former gives rise mainly to technical difficulties,
arising from the impossibility of using such a powerful tool as Fourier
transforms. The latter involves some deeper questions concerning the 
role of large scale geometry in phase transitions
on important real structures such as amorphous solids, glasses, polymers
and fractals.

Recently, these problems have been successfully addressed using graph  
theory. A graph, i.e. a network composed of sites and
links connecting nearest-neighboring sites, is the most suitable  
geometrical model to describe an irregular system consisting of 
spins coupled by local interactions. In particular, algebric graph
theory gives very interesting results when dealing with continuous symmetry
models, due to the evidence for deep relations between their critical behavior
and the small eigenvalues spectrum of the Laplacian operator.
                             
One of the most representative models in this class is the $O(n)$ model,  
which describes a classical $n-$dimensional spin vector of fixed length.  
It is not in general exactly solved, but a few analytic results have been  
obtained on lattices. For example the lower critical dimension for spontaneous
symmetry  breaking is known from the Mermin-Wagner theorem and the 
$n \to \infty$ limit, which corresponds to the spherical model, 
can be exactly solved on any lattice. These results
provide the basis for a qualitative understanding of the phase diagram
as well as for the $1/n$ expansion of thermodynamical quantities and, in
particular, of critical exponents. 

As for graphs, while an extension of the Mermin-Wagner theorem to non 
translation invariant structures has been proven \cite{mwg}, 
the $n \to \infty$ result  for $O(n)$ models heavily relies on the invariance
properties of the lattice  and therefore cannot be easily generalized. Indeed
the equivalence
between an infinite components model with local constraints and a model with a
global constraint on the spin is lost as a consequence of the lack of symmetry. 
The extension of this result to a generic discrete 
structure would be an important step in the comprehension of continuous
symmetry models and in particular of their phase diagrams and universality 
classes. In this letter we deal with such a problem.
 
We show that on a graph, although the free energy of $O(n)$ models in the 
$n\to\infty$ limit at a generic temperature is not in general equivalent 
to the corresponding free energy of the spherical model, the singular parts
of the two free energies concide. Therefore on a generic discrete structure
the equivalence is retrieved asymptotically near the critical point and it 
holds for critical exponents,
which can be shown to depend only on the spectral dimension of the network.
This result strongly supports the possibility of a global geometrical
characterization of irregular networks in the critical region, 
affecting all phenomena related to large scales. 
 
The ferromagnetic classical $O(n)$ Heisenberg model is defined by the  
Boltzmann weight $\exp(- \beta H_{n}) $, where
\begin{equation}
H_{n}[{\bf S}]  = - \sum_{<ij>} J_{ij} {\bf S}_i \cdot {\bf S}_j - 
{\bf h}\cdot \sum_{i} {\bf S}_i   ,
\label{oenne}
\end{equation} 
the sum extends to all links of a certain graph ${\cal G}$ with $N$
sites, $J_{ij} > 0$ are ferromagnetic interactions, which may vary
from link to link, and ${\bf S}_i$ is an $n-$dimensional vector of
fixed length normalized by ${\bf S}_i\cdot{\bf S}_i =n$. The
Hamiltonian (\ref{oenne}) is equivalent to
\begin{equation}
H_{n}[{\bf S}]  = {1\over 2} \sum_{i,j} J_{ij} ({\bf S}_i - {\bf S}_j)^2 - 
{\bf h}\cdot \sum_{i} {\bf S}_i   ,
\label{oennem}
\end{equation}
apart from an additive constant. In the following we will use this
form for $H_n$ which simplifies the mathematical proof. 
The free energy per component is defined by
\begin{equation}
f_n(\beta) = -{1\over Nn} ~{1\over \beta} ~\log Z 
\label{oennel}
\end{equation}       
where the partition function reads
\begin{equation}
Z= \int \prod_{i} \delta ({\bf S}_i\cdot{\bf S}_i-n) \,d{\bf S}_i
\,e^{-\beta H}.
\end{equation}
We shall assume the existence of the thermodynamic limit $N\to\infty$
which turns ${\cal G}$ into an infinite graph within a certain class to be
better specified in the sequel.

The classical fundamental result originally due to \cite{kt} applies
when ${\cal G}$ is a regular lattice ({\it e.g.} a hypercubic
lattice); it establishes a rigorous relation between $O(n)$ models in
the $n\to \infty$ limit and the spherical model. The latter is
defined, on the same graph ${\cal G}$ of the $O(n)$ model, by the
Boltzmann weight $e^{-\beta H^S}$ with Gaussian Hamiltonian
\begin{equation}\label{sferi}
H^S= ~{1\over 2} \sum_{<ij>} J_{ij} (\phi_i -\phi_j)^2 - h  \sum_{i} \phi_i  
\end{equation}
and the spherical constraint $\sum_i \phi_i^2= N$, where $\phi_i$ are
real scalar variables. In the thermodynamic limit $N\to\infty$ we are
allowed to replace (\ref{sferi}) by 
\begin{equation}\label{sferico}
H^S= ~{1\over 2} \sum_{<ij>} J_{ij} (\phi_i -\phi_j)^2 + {1\over 2}
m^2 \sum_i (\phi^2_i -1) - h  \sum_{i} \phi_i  
\end{equation}
where the ``mass'' parameter $m^2$ is fixed to be a precise function
of temperature and magnetic field, {\it i.e.}  $m^2=\mu(\beta,h)$, by
the spherical constraint ``on the average''
\begin{equation}\label{constraint}
  \lim_{N\to\infty} {1\over N} \sum_i \langle \phi_i^2 \rangle = 1 
\end{equation}
Since $H^S$ is Gaussian, one finds (here and in the following we
restrict to the case $h=0$)
\begin{equation}\label{gauss}
f^S(m^2) = - \lim_{N\to\infty}{1\over N\beta} ~\log Z^S=
{1\over 2\beta}\log\det(L+m^2)-{m^2\over 2} \;,\quad \langle \phi_i\phi_j
\rangle= {1\over\beta}(L+m^2)^{-1}_{ij}   
\end{equation}
where $L_{ij}= J_i \delta_{ij} - J_{ij}$, with $J_i = \sum_j J_{ij}$,
is a discrete Laplacian
operator on ${\cal G}$. At this stage it is convenient to better
specify the class of infinite graphs and couplings with which we are
concerned: we shall assume that ${\cal G}$ can be naturally embedded
in a finite dimensional Euclidean space and that $J_i$
is uniformely bounded over ${\cal G}$. For our class of graphs a
generic quantity $q_i$, related to a single point $i$, can be averaged
in a unique way over ${\cal G}$ by $ [q]_{{\cal G}} = \lim_{N\to
\infty} 1/N \sum_i q_i$. The measure $|{\cal G}'|$ of a subgraph
${\cal G}'$ is given by the average value of its characteristic
function $\chi^{{\cal G}'}_i$, defined by $\chi^{{\cal G}'}_i=1$ if $i
\in {\cal G}'$ and $\chi_i=0$ otherwise: $|{\cal G}'|= [\chi]_{{\cal
G}}$.  The asymptotic behavior of the model in the massless limit
$m^2\to0$ is related to the spectrum of the Laplacian operator at low
eigenvalues \cite{debole}. In particular we have the following
singular behaviors as $m^2\to0$
\begin{equation}\label{sing}
{\mathrm{sing}\,}[(L+m^2)^{-1}]_{{\cal G}} = {\mathrm{sing}}\lim_{N\to\infty}
 {1\over N} \sum_{i=1}^N (L+m^2)^{-1}_{ii} = C'~(m^2)^{{{\bar d}\over 2} -1}
\end{equation}
which define the spectral dimension $\bar d$ of the graph. Together
with eqs. (\ref{constraint}) (which nows reads 
$[\langle \phi^2 \rangle]_{{\cal G}}=1$) and (\ref{gauss}), 
eq. (\ref{sing}) determines the the asymptotic form of the function
$m^2=\mu(\beta,0)\equiv\mu(\beta)$, that is 
$\mu(\beta)\sim \beta^{2/({\bar d}-2)}$, $\beta\to\infty$ for ${\bar d}<2$ and
$\mu(\beta)\sim (\beta_c-\beta)^{2/({\bar d}-2)}$, $\beta\to\beta_c^-$ for
$2<{\bar d}<4$ and $\mu(\beta)\sim \beta_c-\beta$ for ${\bar d}>4$, where $\beta_c=[L^{-1}]_{{\cal G}}$. In turns this
implies that the free energy of the spherical model (see
eq. (\ref{gauss})) has the asymptotic form near the critical point
\begin{equation}\label{fasym}
  f^S(\mu(\beta)) \simeq f^S(0) + O(\mu(\beta)).
\end{equation}

It is important to observe that for most infinite graphs, if ${\cal
G}'$ is a subgraph of ${\cal G}$ with non-zero measure, $|{\cal
G}'|>0$, the infrared singularities of $[(L+m^2)^{1}]_{{\cal G}}$ and
$[(L+m^2)^{1}]_{{\cal G}'}$ concide and are described by the same
spectral dimension $\bar d$ and the same coefficient $C'$ of the
singular part \cite{bcd}.  We call these graphs ``pure graphs''.  This
is the case for all fractals (such as Sierpinski gasket and carpet,
t-fractals and so on), bundled graphs (comb lattices, brushes, ...)
and many others (e.g. $NT_D$ \cite{dds}).  However some peculiar cases
of macroscopically inhomogeneous networks, obtained for example by
sticking togheter two pure graphs with different spectral dimensions
by a zero-measure set of links, can present different singular
behaviors on different non zero measure subgraphs.  Now, it can be
shown \cite{bcd} that in these cases, which we will call "mixed", the
number of such subgraphs must be finite and the Gaussian free energy
for the whole graph is simply the weighted sum of the Gaussian free
energies of the subgraphs. This reduces the global problem to the
study of a finite number of pure problems. Therefore, in the following
we will restrict to the pure case.

Let us now come back to the equivalence between the $O(\infty)$ model
and the spherical one. The proof of the classical result mentioned
above deeply relies on the translation invariance of lattices with
simple elementary cells whenever the couplings $J_{ij}$ are constant
over the lattice. It holds in the thermodynamic limit and uses a
saddle points technique \cite{kt} that cannot be generalized to a
generic discrete structure. Here, we will follow a more general and
flexible approach to the infinite component limit. In fact
it can be proven that, provided $0<\epsilon<J_{ij}<J_M<\infty$ (bounded
ferromagnetic couplings) \cite{russi}:
\begin{equation}
-{1\over 2n}\log[(m^2 + J_M)] \le f_n - f^S(m^2) \le
K_1 {1\over N} \sum_i \langle \phi_i^2- 1 \rangle^2 + {K_2\over n}
\label{russi}
\end{equation}
where the constants $K_1$ and $K_2$ do not depend on $m^2$, $N$ and
$n$, $f_n$ is the free energy per component for the $O(n)$ model
defined by (\ref{oennel}), $f^S(m^2)$ is that defined in
eq. (\ref{gauss}) and the average in the right has Boltzmann weight
$e^{-\beta H^S}$, with the Hamiltonian (\ref{sferico}) with generic
$m^2$, i.e where $m^2$ is not yet fixed by the average spherical
constraint (\ref{constraint}). Notice that inequality (\ref{russi})
contains as a particular case the hypercubic lattice result.  Indeed
on a translation invariant structure $\langle \phi_i^2 \rangle$ does
not vary from site to site and the spherical constraint
(\ref{constraint}), corresponding to $m^2=\mu(\beta)$, allows the right
hand side of (\ref{russi}) to vanish for $n\to\infty$, implying the
coincidence of the two free energies.

In the case of a generic discrete structure $\langle \phi_i^2 \rangle$
does change from site to site and one would in general need
site--dependent squared masses $m^2_i$ to enforce $\langle \phi_i^2
\rangle=1$ for every $i$ and obtain an analogous coincidence of the two
free energies in the infinite components limit. Thus in the
thermodynamic limit one would have to solve an infinite number of
equations which could not be reduced to a single one due to the lack of
translation invariance, forbidding the equivalence with the spherical
model where only the {\em average} constraint (\ref{constraint}) holds
and is solved through a {\em single} global parametrization
$m^2=\mu(\beta)$.

The key point of our approach is that we are interested in the critical
properties of the two models. Therefore we will require from the beginning 
only the concidence of the {\it singular parts} of the two free energies. This
requirement has two main consequences: the constraint equations are replaced
by ``constraint inequalities" and, above all, the original  models can be
replaced by modified models with the same singular parts of the free
energies, therefore belonging to the same universality classes. 
Then we will show that a particular global choice $m^2 = {\bar \mu}(\beta)$ 
corresponding to the solution of the constraint equation for a rescaled 
spherical model on ${\cal G}$ do satisfy the constraint inequalities and 
therefore  gives the critical behavior of corresponding rescaled $O(n)$ model
in  the $n\to\infty$ limit. Finally, the rescaled models will be shown to
belong to the same universality class of the original ones.

The first step of our proof consists in obtaining the constraint inequalities.

Let us consider the $n\to\infty$ limit of eq. (\ref{russi}), which we
write as
\begin{equation}\label{russi2}
f^S(m^2) \le f_\infty \le  f^S(m^2) + K_1 [\langle \phi^2 -1\rangle^2]_{{\cal G}}
\end{equation}
and suppose we could show that
\begin{equation}\label{nonlin}
 [\langle \phi^2 -1\rangle^2]_{{\cal G}} \le o(\mu (\beta))
\end{equation}
near the critical point of the spherical model. Then comparing with
the asymptotic form (\ref{fasym}) of $f^S(\mu(\beta)$ itself would
immediatly prove that the two free energies have both the same value
at the critical point {\em and} the same singular parts near it.

Now, for the class of infinte graphs under consideration it can be
shown that (\ref{nonlin}) follows from the infinite set of
linearized inequalities
\begin{equation}
 | [\langle\phi^2-1\rangle_{\mu}]_{{\cal G}'} | \le o(\mu (\beta))
\label{lin}
\end{equation}

where the average is taken over every subgraph ${\cal G}'$ with
$|{\cal G}'|>0$.  We shall call (\ref{lin}) the ``constraint
inequalities" by analogy with the usual approach.

As a matter of fact, it is in general impossible that inequalities
(\ref{lin}) are verified by the solution $\mu(\beta)$ of the global
constraint for the standard spherical model on ${\cal G}$.  However,
we prove the following:
\begin{itemize}  
\item it is possible to find a modified set of couplings
$\{J_{ij}'\}$ and site-dependent masses $\{m'^2_i\}$, both functions
of $m^2$, which define two modified models on ${\cal G}$ with
Hamiltonians $H'^S$ and $H'_{n}$ such that the corresponding
constraint inequalities (\ref{lin}) are satisfied by the solution of
the (modified) spherical model on ${\cal G}$, namely $m^2={\bar \mu}(\beta)$,
with ${\bar \mu}(\beta)=O(\mu(\beta)$ near criticality. Therefore,
from the inequality (\ref{russi2}), the two corresponding free
energies $f'_{\infty}(\beta)$ and $f'^S(\bar \mu(\beta))$ have the
same singular part;
   
\item the rescalings $\{J_{ij}\}\to \{J_{ij}'\}$ and
$\{m^2_i\}\to \{m'^2_i\}$ do not
affect the singular parts of the free energies for both the $O(n)$ and the
spherical models and therefore such modified models belong to the same
universality classes as the original ones. 
\end{itemize}    

To prove the first point let us introduce the set of couplings and masses
\begin{equation}\label{jmrisc}
J'_{ij}~= ~J_{ij}\sqrt{[a_i + b_{i}(m^2)] [a_j + b_{j}(m^2)]} \;,
\quad m'^2_i~= m^2 [a_i + b_{i}(m^2)]
\end{equation}
where $0<\epsilon<a_i< A<\infty$ and the $b_i (m^2)$ are functions to
be determined which vanish for $m^2 \to 0$. Such a rescaling can be
seen as a bounded rescaling of the scalar spin variables $\phi_i \to
\phi'_i = \phi_i [a_i + b_{i}(m^2)]$ with the original Hamiltonian
(\ref{sferico}). Therefore, the inequalities (\ref{lin}) for the
modified model $H'^S$ can be written in terms of the $\phi_i'$.  Now
setting $m^2=\mu(\beta)$ the constants $a_i$ and the functions
$b_i(\mu)$ can always be chosen self--consistently to satisfy the
inequalities to $o(\mu(\beta)$. This can be done at the same time for
every subgraph ${\cal G}'$ with $|{\cal G}'|>0$, as required by
eq. (\ref{lin}), because the leading singularity of $[\langle
\phi^2\rangle]_{{\cal G}'}$, including its coefficient, is universal,
that is equal to that of $[\langle \phi^2\rangle]_{{\cal G}}$.

Then for $n\to\infty$ the $O(n)$ model with coupling $J'_{ij}$ 
and the spherical model, defined by the same
set of $J'_{ij}$ satisfy the constraint inequalities.
Therefore, their free energies has the same singular part and the two models
have the same critical properties.

Now, let us consider the universality classes of these models. The 
critical behavior of the spherical model on a graph can be expressed
in term of its spectral dimension $\bar d$, defined by (\ref{sing}).
This has been shown to be invariant under bounded rescaling of couplings
$J_{ij}$ \cite{forte} such as the ones corresponding to the set 
$J'_{ij}$. Therefore, the two spherical models described by the $J_{ij}$ and
$J'_{ij}$ belong to the same universality class.
On the other hand by the generalized Griffith inequalities \cite{jaffe}
the correlation functions of the $O(n)$ model corresponding $H'_n$
has the same critical behavior of the correlation function of the model
$H_n$. Then they belong to the same universality classes.
This completes our proof.

Our result on the equivalence of critical regimes for the spherical
and $O(\infty)$ models on graphs provides a key step 
for the comprehension of phase transitions on general networks.
The spherical critical exponents, which are exactly known and depend
on the spectral dimension of the graph, are the starting point for
the determination of the corresponding ones for $O(n)$ models,
via the $1/n$ expansion. This expansion is a fundamental tool
to analyze the existence of geometrical universality classes on graphs:
indeed if its coefficients can be shown to depend only on the spectral
dimension $\bar d$, then their existence is proved.

%**************************************

\end{document}